\documentclass[preprint,showpacs,amsmath,amssymb]{revtex4-2}

\usepackage{amsmath}
\usepackage{graphicx}
\usepackage{amsfonts}% Include figure files
\usepackage{dcolumn}% Align table columns on decimal point
\usepackage{bm}% bold math
\usepackage{color}

\begin{document}

\title{Revealing Physical Mechanisms of Pattern Formation and Switching in Ecosystems via Nonequilibrium Landscape and Flux}

\author{Jie Su$^{1}$}
\author{Wei Wu$^{1}$}
\author{Denis Patterson$^{2,3,4}$}
\author{Simon Asher Levin$^{2,3}$}
\thanks{Corresponding Author: slevin@princeton.edu}
\author{Jin Wang$^{1,5}$}
\thanks{Corresponding Author: jin.wang.1@stonybrook.edu}

\affiliation{1.Center for Theoretical Interdisciplinary Sciences, Wenzhou Institute, University of Chinese Academy of Sciences, Wenzhou 325001, China}
\affiliation{2.High Meadows Environmental Institute, Princeton University, Princeton, NJ 08544, USA}
\affiliation{3.Department of Ecology and Evolutionary Biology, Princeton University, Princeton, NJ 08544, USA}
\affiliation{4.Department of Mathematical Sciences, Durham University, Durham, UK}
\affiliation{5.Department of Chemistry and of Physics and Astronomy, State University of New York of Stony Brook, Stony Brook, New York 11794, USA}
\date{\today}

\begin{abstract}
Spatial patterns are widely observed in numerous nonequilibrium natural systems, often undergoing complex transitions and bifurcations, thereby exhibiting significant importance in many physical and biological systems such as embryonic development, ecosystem desertification, and turbulence. However, how spatial pattern formation emerges and how the spatial pattern switches are not fully understood. Here, we developed a landscape-flux field theory via the spatial mode expansion method to uncover the underlying physical mechanism of the pattern formation and switching. We identified the landscape and flux field as the driving force for spatial dynamics and applied this theory to the critical transitions between spatial vegetation patterns in semi-arid ecosystems, revealing that the nonequilibrium flux drives the switchings of spatial patterns. We uncovered how the pattern switching emerges through the optimal pathways and how fast this occurs via the speed of pattern switching. Furthermore, both the averaged flux and the entropy production rate exhibit peaks near pattern switching boundaries, revealing dynamical and thermodynamical origins for pattern transitions, and further offering early warning signals for anticipating spatial pattern switching. Our work thus reveals physical mechanisms on spatial pattern-switching in semi-arid ecosystems and, more generally, introduces a useful approach for quantifying spatial pattern switching in nonequilibrium systems, which further offers practical applications such as early warning signals for critical transitions of spatial patterns.
\end{abstract}
%\pacs{05.60.-k,05.40.-a,82.70.Dd}

\maketitle

%\section *{Key words}
%spatial pattern, semi-arid ecosystem, critical transition, mode expansion method, landscape-flux theory, transition path

\section{Introduction}
Spatial patterns are widely observed in numerous nonequilibrium systems across a range of scientific fields, prominent examples include Turing patterns in chemical reactions~\cite{SP-C1,SP-C3,SP-C4}, pattern formation during embryonic development \cite{SP-embryo1,SP-embryo2,SP-embryo3,SP-embryo4}, patterns in vegetation or animal distributions in ecological models~\cite{SP-ecology1,SP-ecology2,SP-ecology3,SP-ecology4,SP-ecology5, SP-book1, SP-book2}, and complex spatial patterns emerging in turbulence~\cite{SP-T1,SP-T2}. These various spatial patterns have attracted considerable interest for their complex emergence and evolution~\cite{SP-ecology6,SP-tumour,EW1,EW2,EW3}. For instance, the complex ecosystems and Earth system components can evade tipping points through the spatial pattern formation, such as Turing patterns and coexistence state~\cite{SP-ecology6}. Specific spatial patterns of proliferation and necrosis can explain clonal expansion, emergence of parallel evolution and microdiversity in tumours~\cite{SP-tumour}. Despite these significant progresses, how the spatial patterns form and switch are not fully understood. Such an understanding requires global description. Here, we developed a landscape and flux field theory via spatial mode expansion method to reveal the physical mechanism of pattern formation and switching. The landscape and flux theory for well mixed spatially homogeneous systems identified the driving forces as landscape gradient/flux and has been successful in investigating global information as well as the dynamical and thermodynamical mechanisms of nonequilibrium systems involving the cell cycle, cell differentiation and development, cancer and ecological systems \cite{flux0,flux1,flux2,flux3,flux4,flux5,flux6,flux7,flux8,flux10,flux11}. We need to generalize this theory to spatially extended systems \cite{flux-SP1,flux-SP2,flux-SP3,flux-SP4}. However, exact solutions for nonlinear systems remain almost impossible due to the complex nature of spatially dependent fields involving huge numbers of degrees of freedom (DOFs) in the spatial locations. Therefore, it is of paramount importance to find a suitable method to describe the global dynamics and evolution of patterns in spatial nonequilibrium systems.

An effective approach to address this problem is to apply the mode expansion method from field theory \cite{mode2,mode3,mode1}, which maps numerous DOFs of spatial locations onto certain representative spatial modes. This mapping can facilitate the transformation of the complex functional probabilistic evolution equation into a simpler Fokker-Plank equation in the mode space. By developing the landscape and flux theory in this mode space, two key benefits for investigating spatial patterns can then be realized. Firstly, we can quantify the potential landscape in the mode space, which reflects the global stability of the spatial pattern systems; such as locations and weights (potential depths) of steady states. Secondly, we can obtain the flux field and the entropy production rate (EPR), which can be used to explore the nonequilibrium dynamical and thermodynamical mechanisms of the emergence and switching of spatial patterns. Therefore, introducing the landscape and flux theory with the mode expansion method to the study of spatially patterned nonequilibrium systems has significant potential to generate deeper physical insights into pattern dynamics and possible practical applications, such as quantifying the likelihood of regime shifts and thereby providing early warning signals.

In this paper, we report a practical example to highlight the potential of applying the nonequilibrium landscape and flux theory combined with the mode expansion method (with appropriate truncations) to spatially extended systems. The model we study concerns spatial vegetation patterns located in semi-arid ecosystem \cite{ecosystem1,ecosystem2,ecosystem3}, which are widely distributed in many regions including parts of Africa \cite{dis0,dis1,dis2,dis3}, Australia \cite{dis4,dis5}, and Mexico \cite{dis6}. The vegetation exhibits diverse spatial patterns, such as stripe, gap, spot \cite{ecosystem1,ecosystem2,ecosystem3,dis0,ecosystem4} and irregular mosaics \cite{dis2,dis3}. It is noteworthy that these spatial patterns may provide an indication of ecosystem resilience, for instance, spot patterns have been shown to be a potential early warning signal of desertification \cite{ecosystem2,D1,D2,D3,D4}. Beyond common factors such as the rainfall, the positive feedback regulation between the plant biomass and water also plays a significant role in generating different types of spatial vegetation patterns located in semi-arid ecosystem~\cite{ecosystem2,ecosystem3,feedback1,feedback2,feedback3}. That is, as the feedback intensity increases, the vegetation pattern switches from gap to stripe and ultimately to spot pattern~\cite{ecosystem3}.

In the current study, we begin with stochastic evolution equations for the vegetation biomass and water fields that are then transformed into reduced stochastic Langevin dynamic equations and a Fokker-Plank equation based on several representative spatial modes by using the mode expansion method with appropriate truncations. Subsequently, by applying the landscape and flux theory in the mode space, we quantify the potential landscape of the semi-arid ecosystems which provides a global picture of the dynamics. We observe that as the soil-water diffusion feedback intensity increases, the landscape transforms successively from gap to gap/stripe, stripe, stripe/spot, and finally spot states. Besides, we find the nonequilibrium flux inside the landscape often acts against the potential gradient, providing a driving force for the system to switch from one spatial pattern to another one. Detailed analysis reveals that the non-overlapping transition paths between alternative stable states, reflecting the time-reversal symmetry breaking of the system, is attributed to the combined effect of the potential gradient force, the flux force and the spatially dependent noise force. Moreover, the switching time between states corresponds to the barrier height as well as the closest basin distance. Both the transition paths and the switching time offer new insights into the physical mechanisms switching between spatial patterns. More interestingly, both the averaged flux and EPR present peaks near the phase boundaries, not only revealing the dynamical and thermodynamical mechanisms of the critical transition, but also providing early warning signals for the desertification. Furthermore, additional simulations explore the influence of ecosystem size (noise) and other parameters, demonstrating the generality and robustness of our method.

\section{Results and Discussion}

\subsection{Model and Methods}

We focus on the well-known semi-arid model of vegetation biomass and water~\cite{ecosystem1} with the soil-water diffusion feedback~\cite{ecosystem2,ecosystem3} (see SI for details). By adding noise terms to the deterministic evolution equation Eq.S3, the stochastic partial differential equations (PDEs) of semi-arid ecosystem model can be written as follows:
\begin{equation}
\begin{aligned}
\frac{\partial{n}}{\partial{t}} &= wn^2 - mn + \Delta n + \zeta_n(t), \\
\frac{\partial{w}}{\partial{t}} &= a - w - wn^2 + \alpha\Delta(w-\beta n) + \zeta_w(t).
\label{eq:Model3}
\end{aligned}
\end{equation}
\noindent Herein, $w$ and $n$ represent the water and vegetation biomass, respectively; $a$ controls water input; $m$ measures plant losses; $\alpha$ represents the diffusion rate of water; and $\beta$ denotes the water-uptake ability of the roots, i.e., the soil-water diffusion feedback intensity. $\zeta_n$ and $\zeta_w$ are stochastic terms with the time correlation satisfying with \cite{flux-SP1,FFPE1} $\langle\boldsymbol{\zeta}(\mathbf{r},t)\boldsymbol{\zeta}(\mathbf{r}',t')\rangle = \delta(t-t')\times \mathbf{M}$, where $\boldsymbol{\zeta}$ is a vector consisting of the components of $\zeta_n$ and $\zeta_w$, $\mathbf{M}$ is a $2\times2$ matrix:
$$
\mathbf{M} = 
\begin{bmatrix}
(wn^2 + mn)\delta(\mathbf{r}-\mathbf{r}^\prime)  & -wn^2\delta(\mathbf{r}-\mathbf{r}^\prime) \\
+2\nabla\cdot\nabla^\prime [n\delta(\mathbf{r}-\mathbf{r}^\prime)] & \\
  & (a + w + wn^2)\delta(\mathbf{r}-\mathbf{r}^\prime) \\
-wn^2\delta(\mathbf{r}-\mathbf{r}^\prime)  & +2\alpha\nabla\cdot\nabla^\prime[(w-\beta n)\delta(\mathbf{r}-\mathbf{r}^\prime)] \\
\end{bmatrix}.
$$
We fix $m=1.85$, $a=8.0$, $\alpha=50.0$ with reflecting boundary conditions. The parameters used in this study are derived and estimated from the published data in Ref.\cite{ecosystem3} (Details on the parameter selections can be found in the supporting information, SI). It is found that the vegetation pattern undergoes a progression from gap to stripe and ultimately spot shapes as $\beta$ increases (see details in SI and Fig.S1). These results demonstrate that $\beta$ plays a very important role during the pattern formation and switching process in semi-arid ecosystems.

Based on the stochastic PDEs (Eq.\ref{eq:Model3}), the corresponding probability evolution is predictable and can be described as the functional Fokker-Planck equation (FPE):
\begin{equation}
\begin{aligned}
\frac{\partial P}{\partial t} &= -\int_0^{L_b}dx\int_0^{L_b}dy\{ \frac{\delta}{\delta n(x,y)}A_n + \frac{\delta}{\delta w(x,y)}A_w \}P + \\
& \frac{1}{2}\int_0^{L_b}dx\int_0^{L_b}dy\int_0^{L_b}dx^\prime\int_0^{L_b}dy^\prime \{\frac{\delta^2}{\delta n(x,y)\delta n(x^\prime,y^\prime)}B_{nn} + \\
& \frac{\delta^2}{\delta w(x,y)\delta w(x^\prime,y^\prime)}B_{ww} + 2\frac{\delta^2}{\delta n(x,y)\delta w(x^\prime,y^\prime)}B_{nw}  \}P.
\label{eq:FFPE}
\end{aligned}
\end{equation}
\noindent Herein, $A$ and $B$ are respectively the ``convective" and ``diffusion" parts, with specific expressions described as Eq.S6 in SI. However, exact solutions for Eq.\ref{eq:FFPE} remain almost impossible due to its huge numbers of DOFs in the spatial locations.

To address this, we introduce the mode expansion method~\cite{mode2,mode3,mode1} to map these huge numbers of DOFs into spatial modes. Here, we perform a Fourier transformation and use the Fourier cosine transform to expand $n$ and $w$ in the spatial mode space: $n(x,y,t) = \sum_{i=0}^{+\infty}\sum_{j=0}^{+\infty} N_{ij}(t)\cos\left((ix+jy)k\pi/L_b\right)$ and $w(x,y,t) = \sum_{i=0}^{+\infty}\sum_{j=0}^{+\infty} W_{ij}(t)\cos\left((ix+jy)k\pi/L_b\right)$ (see details and explanations in SI). Hence, the functional FPE (Eq.\ref{eq:FFPE}) in the real space can be changed to the FPE in the spatial mode space based on the spatial modes ($N_{ij}$, $W_{ij}$) (see details and definitions of $\bar{A}$, $\bar{B}$ in SI):
\begin{equation}
\begin{aligned}
\frac{\partial P}{\partial t} = & -\sum_{i=0}^{+\infty}\sum_{j=0}^{+\infty} \{ \frac{\partial \bar{A}_{n,ij}}{\partial N_{ij}} + \frac{\partial \bar{A}_{w,ij}}{\partial W_{ij}} \}P + \frac{1}{2L_b^2}  \sum_{i=0}^{+\infty}\sum_{j=0}^{+\infty} \sum_{i^\prime=0}^{+\infty}\sum_{j^\prime=0}^{+\infty}  \\
& \{ \frac{\partial^2 \bar{B}_{nn,ij,i^\prime j^\prime}}{\partial N_{ij}\partial N_{i^\prime j^\prime}} + 2\frac{\partial^2 \bar{B}_{nw,ij,i^\prime j^\prime}}{\partial N_{ij}\partial W_{i^\prime j^\prime}} + \frac{\partial^2 \bar{B}_{ww,ij,i^\prime j^\prime}}{\partial W_{ij}\partial W_{i^\prime j^\prime}} \}P. \\
\label{eq:FFPE2}
\end{aligned}
\end{equation}

Then we set appropriate truncations~\cite{mode2,mode3,mode1} to select several key spatial modes, including $N_{00}$, $N_{01}$, $N_{10}$ and $N_{11}$ (components of $\mathbf{N}$) as well as $W_{00}$, $W_{01}$, $W_{10}$ and $W_{11}$ (components of $\mathbf{W}$) (see SI for details). The rationale of the truncation is to keep only the low lying or low energy modes and cut off all high energy excited modes, and we make sure that all kinds of Turing patterns including stripe or hexagonal even uniform shapes can be reproduced by these selected modes with truncations completely and easily. The evolution of these truncated modes can be described as the following equations:
\begin{equation}
\dot{\mathbf{N}} = \mathbf{G}(\mathbf{N},\mathbf{W}), \dot{\mathbf{W}} = \mathbf{H}(\mathbf{N},\mathbf{W}).\\
\label{eq:N_W}
\end{equation}
\noindent Herein, $\mathbf{G}$ and $\mathbf{H}$ consist of functions $G_{N_{00}}$, $G_{N_{01}}$, $G_{N_{10}}$, $G_{N_{11}}$, and $H_{W_{00}}$, $H_{W_{01}}$, $H_{W_{10}}$, $H_{W_{11}}$ respectively, and their specific expressions are described as Eq.S10-S17 in SI. We then perform simulations based on these dynamic equations with $L_b=80$ and $k=23$ (see details in SI). As a result, we find that as $\beta$ increases, the vegetation pattern changes from uniform to gap, followed by stripe and finally spot pattern (see phase diagram and typical snapshots in SI and Fig.S2), consistent with the results obtained from simulations in the real space (Fig.S1).

Moreover, the time scales of relaxation for the diverse modes differ significantly. In Fig.S3, it is evident that $N_{01}$, $N_{10}$ and $N_{11}$ exhibit slower relaxation compared to the other five modes. Thus, after employing an adiabatic approximation, the other five modes can be expressed in terms of $N_{01}$, $N_{10}$ and $N_{11}$, meaning that the dynamic equations only depend on these three slow-changing modes. Therefore, Eq.\ref{eq:FFPE2} can be simplified to the Fokker-Plank equation based on $N_{01}$, $N_{10}$ and $N_{11}$ (see details in SI):
\begin{equation}
\frac{\partial P}{\partial t} = -\sum_{ij=01,10,11}\frac{\partial \dot{N}_{ij} }{\partial N_{ij}} P + \frac{1}{2L_b^2} \sum_{ij=01,10,11}^{i^\prime j^\prime=01,10,11}\frac{\partial^2 D_{ij,i^\prime j^\prime}}{\partial N_{ij}\partial N_{i^\prime j^\prime}} P. \\
\label{eq:FPE}
\end{equation}
\noindent Herein, $D_{ij,i'j'}$ is the element of the diffusion matrix with the expression described as Eq.S22-S27 in SI. 

According to Eq.\ref{eq:FPE}, the stochastic Langevin equations in the mode space can read as (see parameters and details in SI):
\begin{equation}
\begin{aligned}
\frac{d N_{01}}{d t} &= F_{N_{01}}(N_{01},N_{10},N_{11}) + \xi_{N_{01}}(t),\\
\frac{d N_{10}}{d t} &= F_{N_{10}}(N_{01},N_{10},N_{11}) + \xi_{N_{10}}(t),\\
\frac{d N_{11}}{d t} &= F_{N_{11}}(N_{01},N_{10},N_{11}) + \xi_{N_{11}}(t).
\label{eq:LE}
\end{aligned}
\end{equation}
\noindent Then we can apply the landscape and flux theory \cite{flux0,flux1,flux2,flux3,flux4,flux5,flux6,flux7,flux8,flux10,flux11} in the mode space, and the effective potential $U$ and the steady-state nonequilibrium flux $\mathbf{J}_{ss}$ can be defined as (see details and parameter definitions in SI):
\begin{equation}
U(N_{01},N_{10},N_{11})=-\ln P_{ss}(N_{01},N_{10},N_{11}),
\label{eq:Potential}
\end{equation}
\begin{equation}
\mathbf{J}_{ss}=\mathbf{F}P_{ss}-\nabla\cdot (\mathbf{D}P_{ss}).
\label{eq:Flux}
\end{equation}

\noindent Different from the case in equilibrium that the driving force is only related to the potential gradient, i.e., the flux is equal to $0$, In the steady state of the nonequilibrium system, due to the non-zero rotational flux field, the driving force ($\mathbf{F}$) can be decomposed into the gradient part ($\mathbf{F}_{gradient}$), the curl part ($\mathbf{F}_{curl}$) and the part related to the spatial dependent noise ($\mathbf{F}_D$):
\begin{equation}
\mathbf{F} =\mathbf{F}_{gradient}+\mathbf{F}_{curl}+\mathbf{F}_{D}=-\mathbf{D}\cdot\nabla U+\mathbf{J}_{ss}/P_{ss}+\nabla\cdot\mathbf{D}.
\label{eq:Force}
\end{equation}

\subsection{The global stability of semi-arid ecosystems}

\begin{figure*}
\centering
\includegraphics[width=0.9\columnwidth]{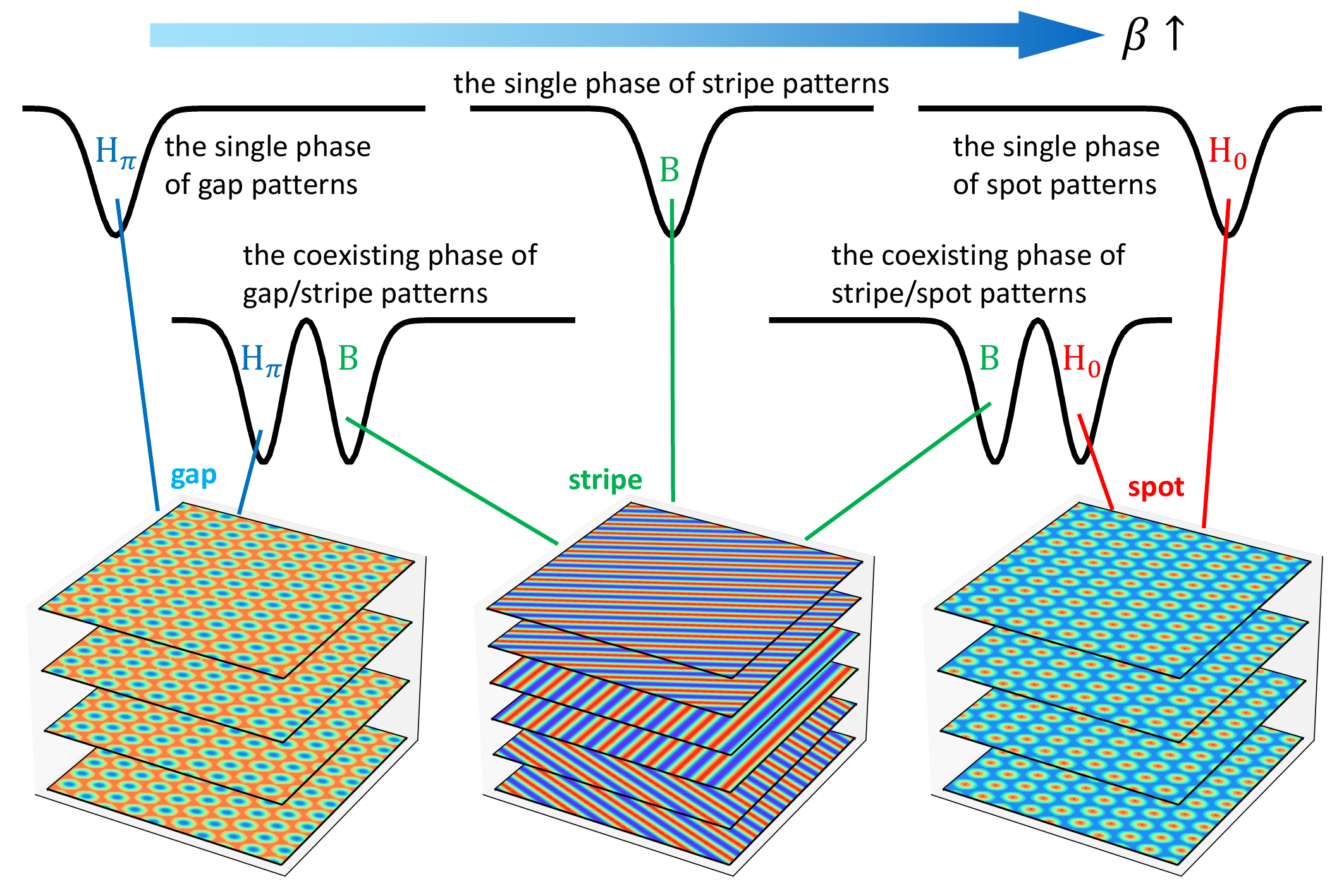}
\caption{\textbf{The conceptual picture of the potential landscape depending on $\beta$.} As $\beta$ increases, the landscape undergoes transitions: from a single potential landscape basin in $H_\pi$ (gap pattern) state to two basins in both $H_\pi$ and $B$ states, then to a single basin in $B$ (stripe pattern) state, followed by two basins in both $B$ and $H_0$ states and finally to a single basin in $H_0$ (spot pattern) state. Due to different phase shifts or orientations among spatial patterns, the number of $H_\pi$, $B$ and $H_0$ states are $4$, $6$ and $4$, respectively.}
\label{fig:result2}
\end{figure*}

Firstly, we present the conceptual picture of the quantified potential landscape depending on the soil-water diffusion feedback intensity $\beta$, as illustrated in Fig.\ref{fig:result2}. As $\beta$ increases, the schematic diagram of the landscape in the mode space shifts from a single potential basin in $H_\pi$ (gap pattern) state to two basins in both $H_\pi$ and $B$ states, then to a single basin in $B$ (stripe pattern) state, followed by two basins in both $B$ and $H_0$ states and finally to a single basin in $H_0$ (spot pattern) state. $H_\pi$, $B$ and $H_0$ states respectively have $4$, $6$ and $4$ steady states located at different regions in the mode space (see details in SI and Fig.S4). These steady states are degenerate with equal probability or potential on the landscape, and patterns located in them all have the same shape but exhibit different phase shifts or orientations between them (see detailed description in SI and Fig.S5-S7). 

\begin{figure*}
\centering
\includegraphics[width=1.0\columnwidth]{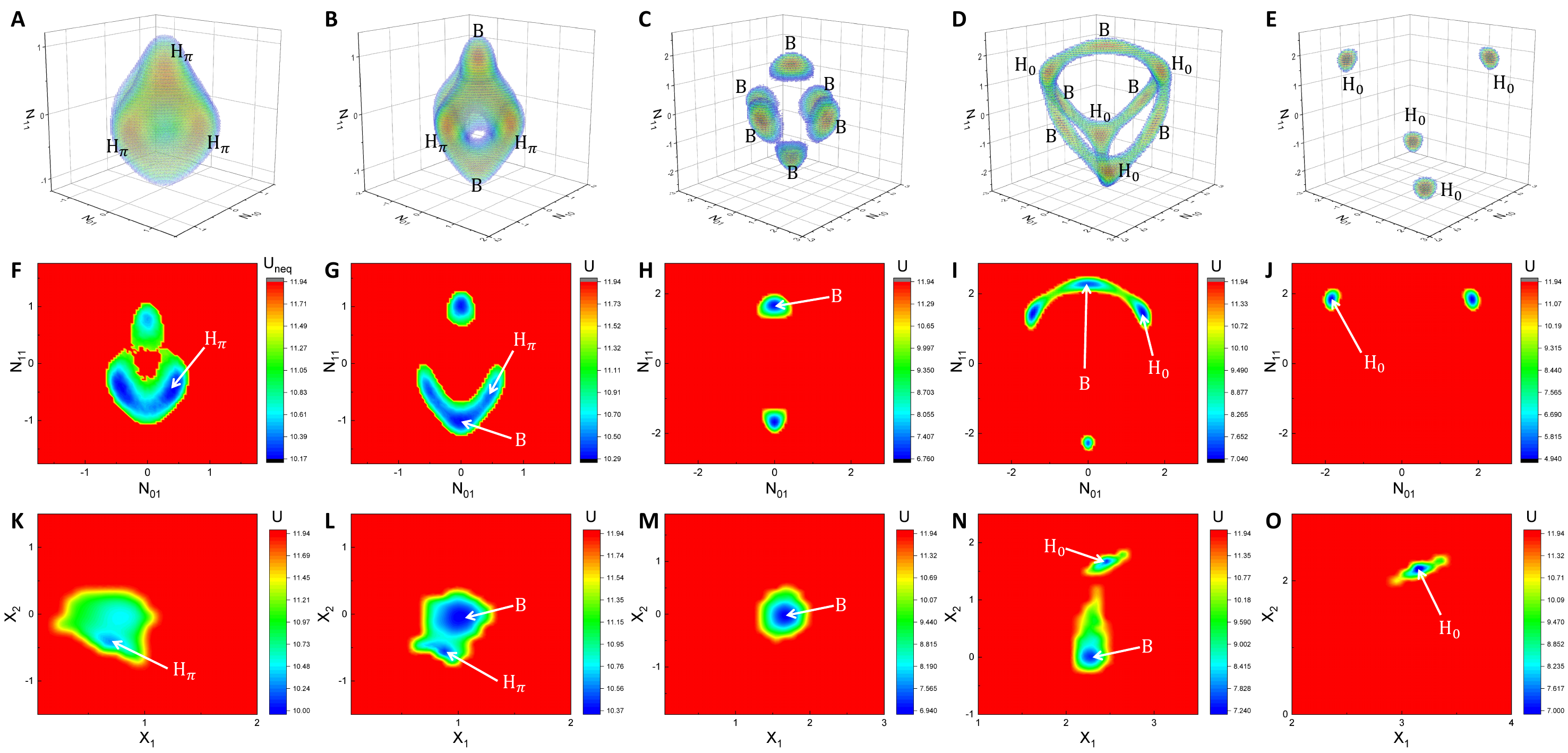}
\caption{\textbf{The typical potential landscape of the semi-arid ecosystem in the mode space.} Landscape in the 3D mode space (\textbf{A}-\textbf{E}), the sections of $N_{01}-N_{10}=0$ (\textbf{F}-\textbf{J}), and the $X_1$-$X_2$ mapping space (\textbf{K}-\textbf{O}). The parameters are $\beta=0.003$ for (\textbf{A},\textbf{F},\textbf{K}), $0.004$ for (\textbf{B},\textbf{G},\textbf{L}), $0.009$ for (\textbf{C},\textbf{H},\textbf{M}), $0.018$ for (\textbf{D},\textbf{I},\textbf{N}), and $0.026$ for (\textbf{E},\textbf{J},\textbf{O}). The color in (\textbf{A}-\textbf{E}) changes from blue to red representing that the potential changes from high to low.}
\label{fig:result3}
\end{figure*}

Then simulations of stochastic Langevin equations (Eq.\ref{eq:LE}) are performed and the potential landscape based on the mode space can be calculated by Eq.\ref{eq:Potential}. The obtained 3-dimensional landscape of typical patterns are illustrated in Fig.\ref{fig:result3}A-E; 2-dimensional sections of the plane $N_{01}-N_{10}=0$ are presented in Fig.\ref{fig:result3}F-J. For $\beta$ equal to $0.003$, $0.004$, $0.009$, $0.018$ and $0.026$, the potential landscape in the section of $N_{01}-N_{10}=0$ presents pairs of potential basins located in $H_\pi$ state (Fig.\ref{fig:result3}F), $H_\pi$+$B$ states (Fig.\ref{fig:result3}G), $B$ state (Fig.\ref{fig:result3}H), $B$+$H_0$ states (Fig.\ref{fig:result3}I) and $H_0$ state (Fig.\ref{fig:result3}J), respectively, corresponding to the five landscapes shown in the conceptual picture in Fig.\ref{fig:result2}. Furthermore, we introduce a mapping method to merge those degenerate potential basins that are located in the same state together. We fix $X_1=\sqrt{N_{01}^2+N_{10}^2+N_{11}^2}$, $X_2=\text{sgn}({N_{01}N_{10}N_{11}})(\vert N_{01}\vert+\vert N_{10}\vert+\vert N_{11}\vert-X_1)$ with $\text{sgn}(x)$ the sign function, and then map the 3-dimensional mode-space landscape into the $X_1$-$X_2$ phase space. On this basis, potential basins in $H_\pi$, $B$ and $H_0$ states will be mapped into the basin located in the region with $X_1>0$, $X_2<0$; $X_1>0$, $X_2=0$; and $X_1>0$, $X_2>0$; respectively. As shown in Fig.\ref{fig:result3}K-O, the potential basins are quantified and illustrated through the locations in gap, gap/stripe, stripe, stripe/spot and spot states, as $\beta$ increases.

\begin{figure}
\centering
\includegraphics[width=0.9\columnwidth]{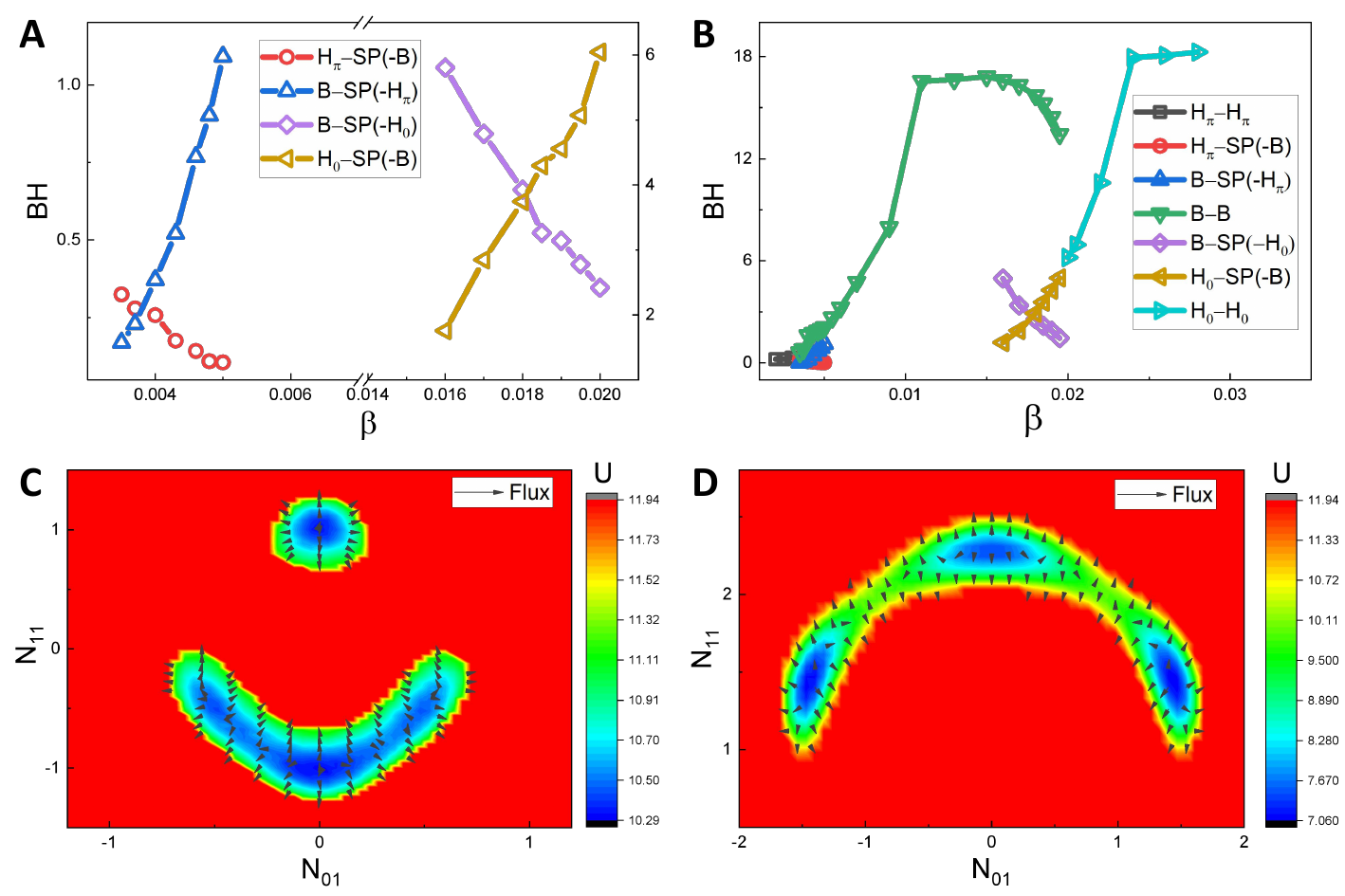}
\caption{\textbf{The barrier height and the flux field of the semi-arid ecosystem.} (\textbf{A}) Dependence of the inter-state barrier height on $\beta$ in the coexisting phases. (\textbf{B}) Dependence of the barrier height on $\beta$, including both the inner-state and inter-state barriers. (\textbf{C},\textbf{D}) The potential section of $N_{01}-N_{10}=0$ (colored background) and flux field (black arrows) of the ecosystem in the coexisting phases with (\textbf{C}) $\beta=0.004$ and (\textbf{D}) $\beta=0.018$.}
\label{fig:result4}
\end{figure}

Based on the landscape, global topography such as the barrier height can be obtained. As shown in Fig.\ref{fig:result4}A, we exhibit the inter-state barrier height (IntBH) in the $X_1-X_2$ phase space and observe two different types of barriers in the coexisting phases. Interestingly, for the gap/stripe coexisting phase, the IntBH between potential basins in the gap state and the saddle point decreases as $\beta$ increases, while the one in the stripe state increases, indicating that the weight of gap and stripe states become lighter and heavier, respectively. Similarly, for the stripe/spot coexisting phase, the IntBH between the basin in the former(latter) state and the saddle point decreases(increases) as $\beta$ increases, implying that the weight of the former(latter) state becomes lighter and lighter(heavier and heavier). In addition to the inter-state barrier between potential basins located in different states, the inner-state barrier between the closest basins situated in the same state can be presented through the landscape in the mode space. As shown in Fig.\ref{fig:result4}B, we observe that the inner-state barrier height (InnBH) in the single phase increases as a whole with the rise of $\beta$. Therefore, the barrier height can be used as a global stability measure quantifying the capability of spatial pattern switching. Higher barrier implies harder for state to switch and thus global stability, while lower barrier indicates higher chances of spatial pattern switching and thus lower stability. Since the spot vegetation pattern implies the desertification precursor, the barrier height of the spot spatial pattern then becomes a corresponding early warning for the environmental protection.

Next, we focus on the nonequilibrium flux from Eq.\ref{eq:Flux}. The obtained flux field for typical patterns in the 3-dimensional mode space are shown in Fig.S8, which is also challenging to see clearly, similar to the potential landscape in the 3-dimensional mode space. Similarly, we choose the section of the plane $N_{01}-N_{10}=0$ to present the flux field more intuitively. As examples in the coexisting phases, two typical landscapes with flux for $\beta=0.004$ and $\beta=0.018$ are shown in Fig.\ref{fig:result4}C and D, respectively. The flux typically points to the opposite direction of the potential gradient, acting as a driving force for the system to switch from one state to the other one. This means that the vegetation pattern can switch more easily between the two pattern states in the coexisting phase with the help of the nonequilibrium flux.

\subsection{The transition paths and rate between different spatial patterns}

\begin{figure}
\centering
\includegraphics[width=0.9\columnwidth]{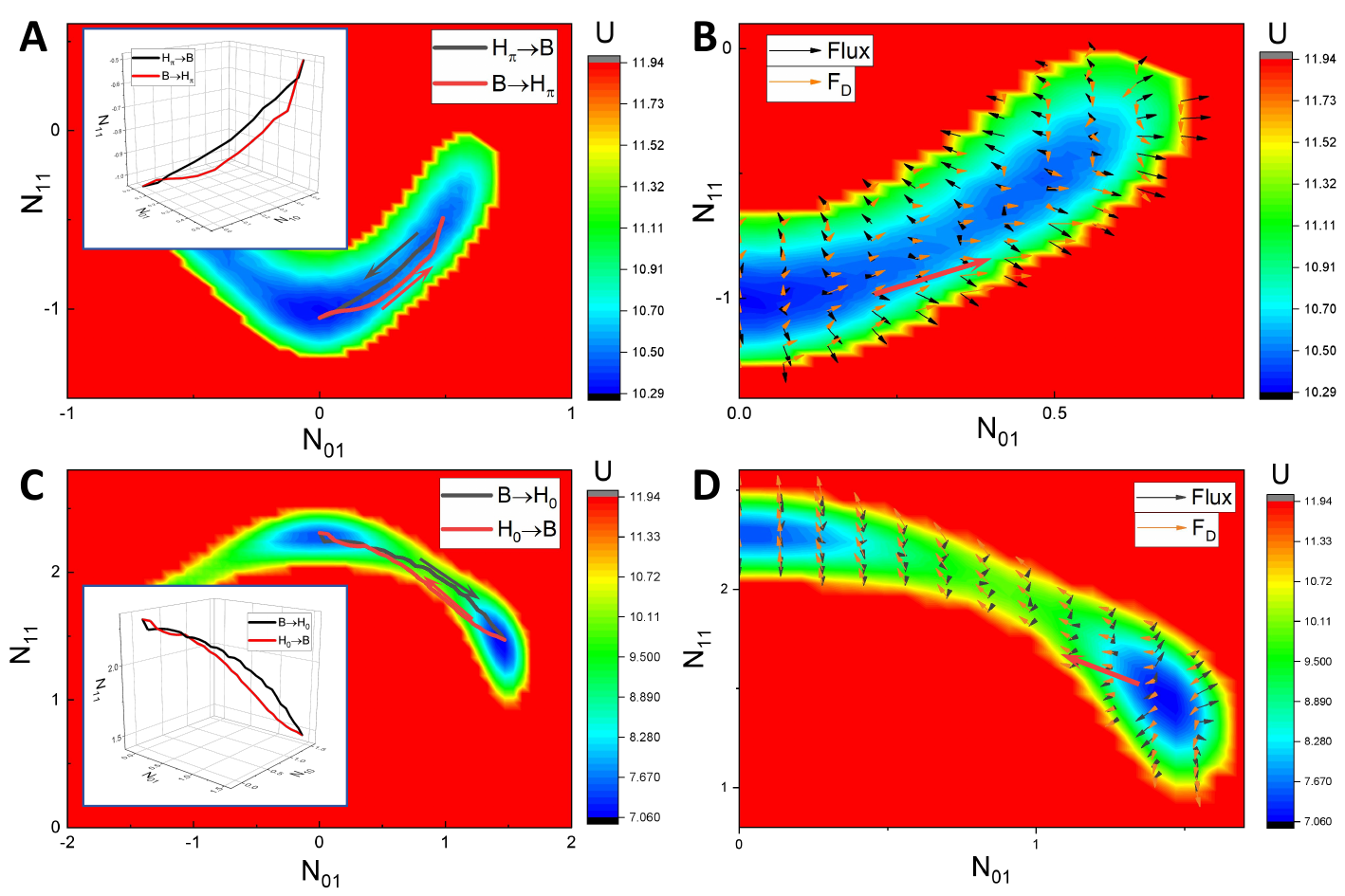}
\caption{\textbf{The transition paths between different states in the coexisting phase.} (\textbf{A}) The transition paths between the gap ($H_{\pi}$) and stripe ($B$) states in the potential section of $N_{01}-N_{10}=0$ for $\beta=0.004$. (\textbf{B}) The zoomed image around the transition paths in (\textbf{A}) consisting of the potential (colored background), flux (black arrows) and the spatial dependent noise focre (orange arrows). (\textbf{C}) The transition paths between the spot ($H_0$) and stripe ($B$) states in the potential section of $N_{01}-N_{10}=0$ for $\beta=0.018$. (\textbf{D}) The zoomed image around the transition paths in (\textbf{C}). The insets in (\textbf{A}) and (\textbf{C}) are the transition paths in the 3D mode space.}
\label{fig:result5}
\end{figure}

In addition to the global information obtained by establishing the landscape and flux field in the mode space, we can also explore how the spatial pattern switching process is realized and how fast this occurs. A pair of transition paths between the gap state (located in the region with $N_{01}=N_{10}=-N_{11}>0$) and the stripe state (located in the region with $N_{01}=N_{10}$=0 and $N_{11}<0$) for $\beta=0.004$ in the gap/stripe coexisting phase is shown in Fig.\ref{fig:result5}A by using the dominant path approach \cite{path1}. We can observe that the transition path from the gap state to the stripe state (the black line) does not overlap with its inverse path from the stripe state to the gap state (the red line), implying that the time-reversal symmetry is broken in such situations. To explain the non-overlap of this pair of transition paths, a zoomed image around these paths is presented in Fig.\ref{fig:result5}B. It is found that the transition path from the gap state to the stripe state (the black line in Fig.\ref{fig:result5}A) seems to move along with the lowest potential which is easily and naturally to be understood. Interestingly, the transition path from the stripe state to the gap state (the red line in Fig.\ref{fig:result5}A) shifts slightly to the right of the black path, consistent with the combined effect including the potential gradient force, the flux force (black arrows in Fig.\ref{fig:result5}B) and the force related to the spatial dependent noise (orange arrows).

Similarly, for the situation in the stripe/spot coexisting phase at $\beta=0.018$ (Fig.\ref{fig:result5}C), we find that the transition path from the stripe state to the spot state (the black line) moves along with the lowest potential, while the one from the spot state to the stripe state (the red line) shifts slightly to the left of the black path. This can also be explained by the combined effect consisting of the potential gradient force, the flux force and the force related to the spatial dependent noise (Fig.\ref{fig:result5}D). Additionally, the pair of transition paths at $\beta=0.018$ overlap less than those in the case $\beta=0.004$, indicating that the degree of the time-reversal symmetry breaking in the stripe/spot coexisting phase is greater than that in the gap/stripe coexisting phase. Moreover, we focus on the dominant modes on this transition path (the black line in Fig.\ref{fig:result5}C) at $\beta=0.018$ during the nucleation process (discussed in the next section). We find that $N_{01}$ and $N_{10}$ play dominant roles on this path. This indicates that when the ecosystem changes from the stripe pattern to the spot one, it tends to increase the maximum magnitude of vegetation biomass $n$ first (before the nucleation formation) and then change the pattern shape later (see details in SI and Fig.S9,S10). In other words, the emergence of the non-uniform magnitude of vegetation biomass in the stripe pattern predicts the pattern switching to a spot shape, which may herald the onset of desertification.

\begin{figure}
\centering
\includegraphics[width=1.0\columnwidth]{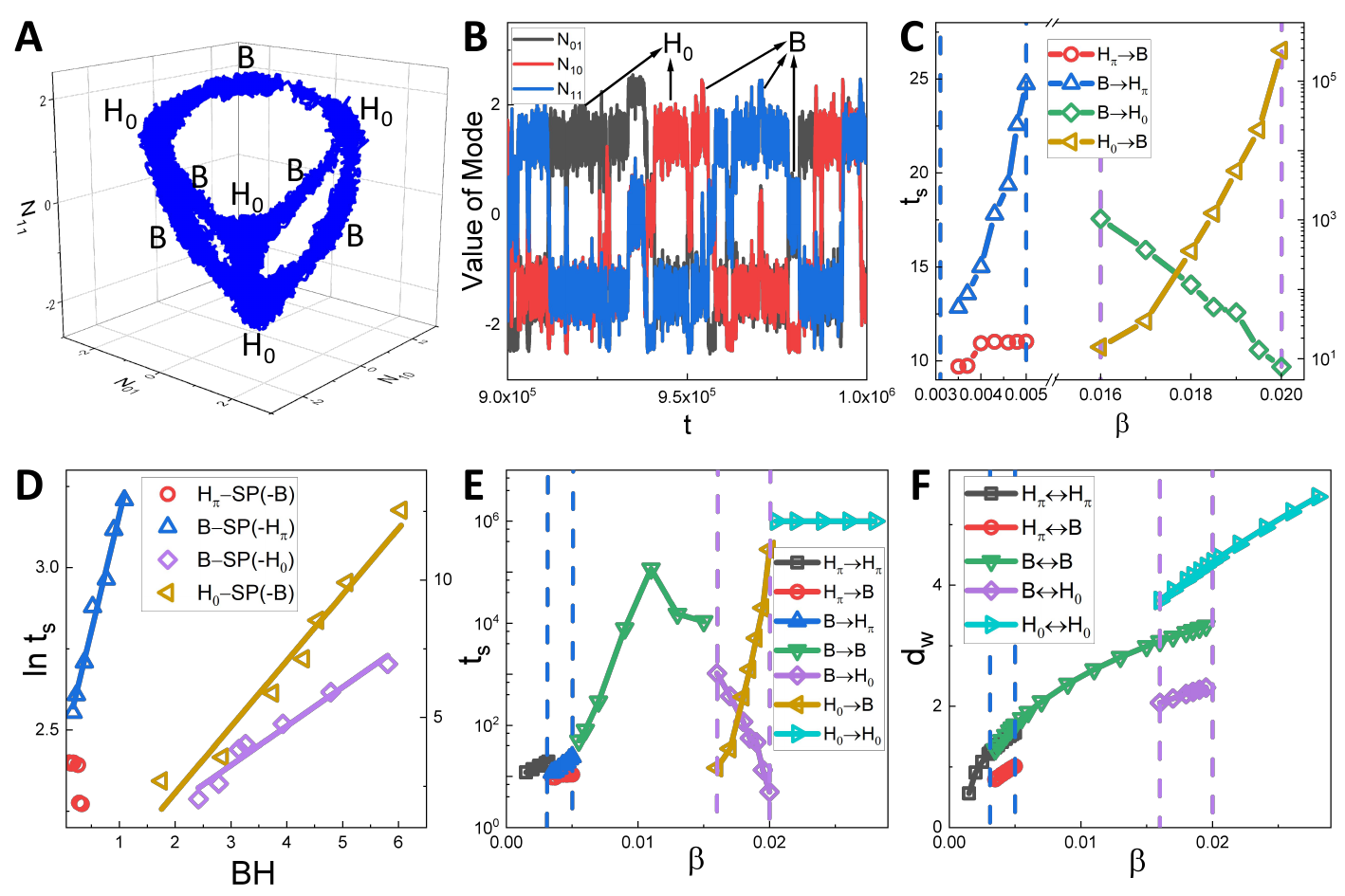}
\caption{\textbf{The transition rate between different states.} (\textbf{A}) Trajectory of the semi-arid ecosystem with $\beta=0.018$ in the steady state. (\textbf{B}) Time series of $N_{01}$, $N_{10}$ and $N_{11}$ in (\textbf{A}). (\textbf{C}) Dependence of the inter-state switching time $t_s$ on $\beta$ in the coexisting phases. (\textbf{D}) $\ln t_s$ as functions of the inter-state barrier height. (\textbf{E}) Dependence of $t_s$ on $\beta$, including both the inter-state and inner-state switching time. (\textbf{F}) Dependence of $d_w$ on $\beta$. The blue and purple dashed lines in (\textbf{C},\textbf{E},\textbf{F}) denote the phase boundaries.}
\label{fig:result6}
\end{figure}

Simultaneously, the transition rate can be understood by investigating the switching time from one state to the another. We perform an additional simulation for one ecosystem evolving in the mode space after reaching the steady state. As an example illustrated in Fig.\ref{fig:result6}A,B, we present the trajectory in 3D mode space and the time series of $N_{01}$, $N_{10}$ and $N_{11}$ at $\beta=0.018$. The time before the system jumps from one state to another is referred to as the single switching time. By performing time averaging, the obtained inter-state switching time $t_s$ dependent on $\beta$ in the coexisting phases are shown in Fig.\ref{fig:result6}C. For the stripe/spot coexisting phase, $t_s$ from the potential basin in the stripe(spot) state to the one in the spot(stripe) state decreases(increases), as $\beta$ increases. All these results are aligned with the behaviors of the IntBH (Fig.\ref{fig:result4}A) and can be similarly understood through it. In other words, the switching time between the states is exponentially related to the barrier height in between. As shown in Fig.\ref{fig:result6}D, the quantity $\ln t_s$ increases linearly as the barrier height increases in the stripe/spot coexisting phase (purple and yellow symbols and lines). Therefore, the speed of spatial pattern switching is strongly related to the barrier height between these states.

However, for the gap/stripe coexisting phase (Fig.\ref{fig:result6}C), both of $t_s$s increase as $\beta$ increases. As illustrated in Fig.\ref{fig:result6}D, $\ln t_s$ switching from the stripe state to the gap state (blue symbols and line) increases linearly as the corresponding barrier height increases, but $\ln t_s$ switching from the gap state to the stripe state (red symbols) doesn't depend on the corresponding barrier height linearly, which cannot be explained solely by the barrier height. IntBH between potential basins in the gap state and the saddle point decreases very slowly as $\beta$ increases (the red line in Fig.\ref{fig:result4}A,B), suggesting that IntBH does not solely determine $t_s$ in such situation. Apart from the barrier height, the distance ($d_w$) between the closest potential basins also affects $t_s$, particularly when the barrier height remains nearly unchanged. As shown in Fig.\ref{fig:result6}F, it is found that all $d_w$s increase as $\beta$ increases. Consequently, the increase of $d_w$ naturally leads to an increase of $t_s$ from wells in the gap state to the ones in the stripe state. Furthermore, we plot $t_s$ of both the inner-state and inter-state switching time in Fig.\ref{fig:result6}E. We observe that as $\beta$ increases, $t_s$ mainly increases in the single phase of gap and stripe states and maintains a large value in the spot state, aligning with the behavior of InnBH as shown in Fig.\ref{fig:result4}B. Therefore, the larger $\beta$ leads to a more stable vegetation pattern, i.e., the ecosystem tends to maintain its configuration rather than change the orientation or phase shift. Additionally, the speed of spatial pattern switching not only depends on barrier height, but also on distance between basins. 

\subsection{Dynamical and thermodynamical mechanisms of the critical transition}

Now we focus on the dynamical and thermodynamical mechanisms of the critical transition in the semi-arid ecosystem, which undergoes a transition from one spatial pattern to another as $\beta$ increases. To capture the dynamics of the ecosystem during the transition, we introduce the averaged flux $J_{ave}$ \cite{flux3,flux11}, motivated by the way in which the nonequilibrium flux (decomposed from the driving force) can effectively characterize the nonequilibrium aspects of the dynamics. To understand the thermodynamics of the transition, we introduce the noise-averaged global entropy production rate (EPR) $e_{p}$  \cite{seifert2012stochastic}. The EPR serves to quantify the degree of detailed balance breaking and time-reversal symmetry in the mode space, or in other words, the extent to which the system is out of equilibrium (see SI for further details).  $J_{ave}$ and $e_p$ may be written in terms of the steady-state flux $\mathbf{J}_{ss}$ as follows:

\begin{equation}
J_{ave} = \langle \vert \mathbf{J}_{ss}(N_{01},N_{10},N_{11}) \vert \rangle,
\label{eq:Jave}
\end{equation}
\begin{equation}
e_{p} = \iiint dN_{01}dN_{10}dN_{11}\frac{\mathbf{J}_{ss}^{\rm T}\cdot\mathbf{D}^{-1}\cdot\mathbf{J}_{ss}}{P_{ss}}.
\label{eq:EPR}
\end{equation}

\begin{figure}
\centering
\includegraphics[width=0.9\columnwidth]{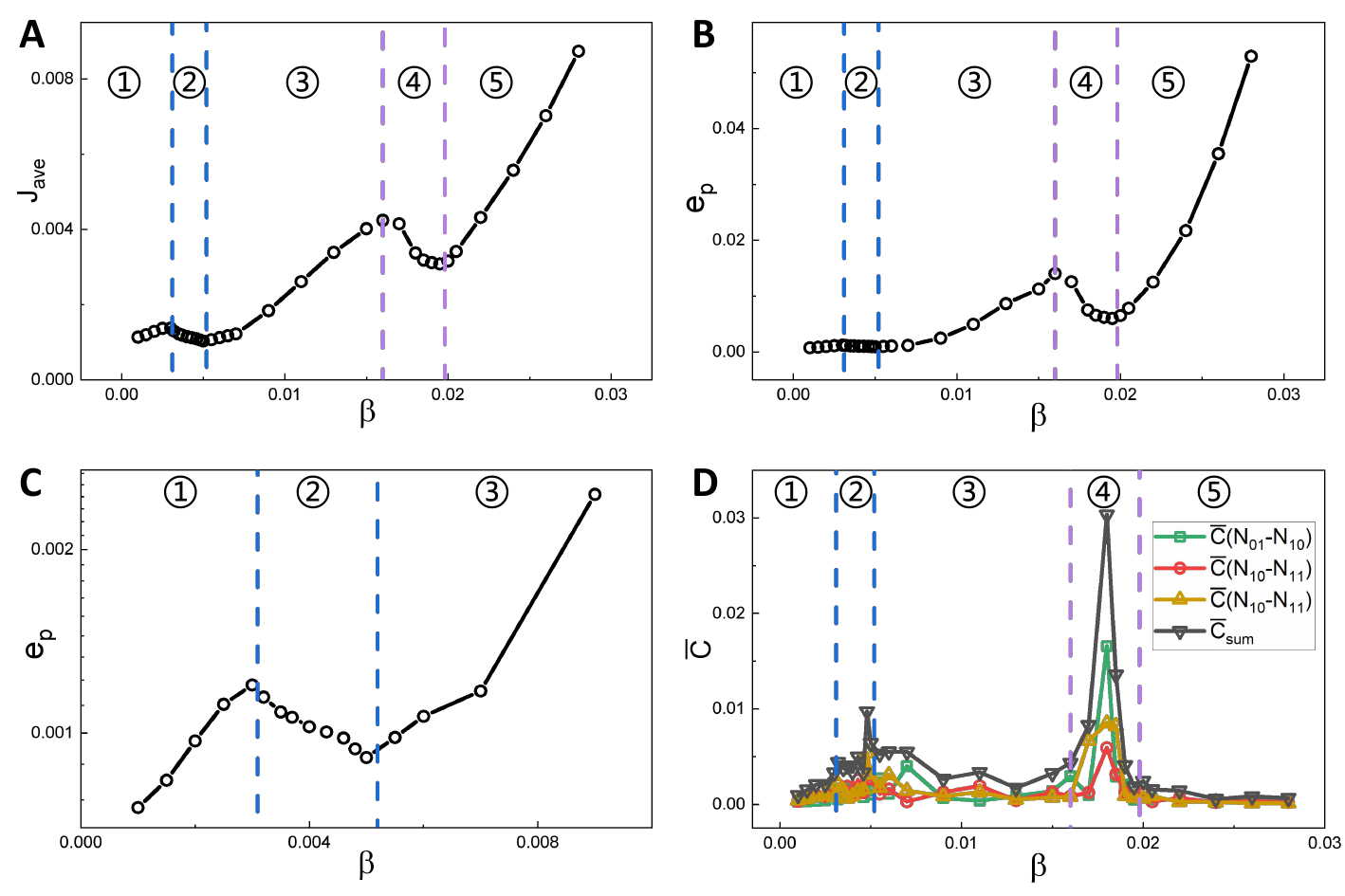}
\caption{\textbf{The dynamical and thermodynamical mechanisms of the critical transition in the semi-arid ecosystem.} (\textbf{A}) Dependence of $J_{ave}$ on $\beta$. (\textbf{B}) Dependence of $e_p$ on $\beta$. (\textbf{C}) The zoomed-in image of (\textbf{B}) at small $\beta$. (\textbf{D}) Dependence of $\overline{C}$ on $\beta$. The regions labeled with $1$-$5$ represent the gap, gap/stripe, stripe, stripe/spot and spot states, respectively. The blue and purple dashed lines denote the phase boundaries. }
\label{fig:result7}
\end{figure}

Fig.\ref{fig:result7}A-C show $J_{ave}$ and $e_p$ as functions of $\beta$. We can observe that $J_{ave}$ increases monotonically as $\beta$ increases in the single phases but decreases in the coexisting phases, resulting in peaks appearing near the phase boundaries. The driving force for the nonequilibrium dynamics can be decomposed into the gradient force of the landscape and the curl force of the nonequilibrium flux \cite{flux1,flux3}. The gradient force due to its convergent nature tends to attract the ecosystem down to the attractor and stabilize it. However, the flux, which breaks the detailed balance due to its rotational nature, is delocalized and favors global movements or switching instead of localizing in one point in the state space. Thus, the dynamical effect of the nonequilibrium flux is to lead to possible instability in one phase while stabilizing the continuous flow between phases. Consequently, the mechanism of the critical transition in the semi-arid ecosystem can be attributed to the emergence of peaks in the averaged nonequilibrium flux. In other words, the nonequilibrium flux serves as the driving force or the dynamical origin for the critical transitions (or bifurcations) and thus the spatial pattern switching.

Similarly, $e_p$ increases monotonically as $\beta$ increases in the single phases and exhibits a general upward trend, indicating that as plant roots become more efficient in local water uptake, the ecosystem dissipates more, leading the system to be further away from equilibrium. Interestingly, $e_p$ also exhibits peaks near the phase boundaries with maxima and minima appearing in the same qualitative pattern as for the averaged flux. Notably, switching from a single phase to a coexisting phase requires significant flux force to transform the one-basin landscape to form the two-basin landscapes, which results in a higher maintenance cost near the beginning boundaries of the critical transition or bifurcation. Thus, the entropy production we calculated in the particular mode space, representing the thermodynamic cost to sustain or vary the states, reflects the degree of detailed balance breaking and time reversal symmetry breaking of the ecosystem. Therefore, the emergence of peaks in nonequilibrium thermodynamic cost characterized by EPR provides an indicator for the thermodynamical mechanism of the critical transition in the semi-arid ecosystem. In other words, the entropy production as nonequilibrium consumption/dissipation serves as the thermodynamic driving force or the thermodynamical origin for the critical transitions or bifurcations and thus the spatial pattern switching. Moreover, since spot vegetation patterns has been shown to be a potential signal of the onset of desertification in previous studies, both the averaged flux and EPR may serve as early warning signals for desertification.

Furthermore, we seek to understand whether the transitions between alternative stable states in this model are ``smooth'' or ``abrupt'' as parameters vary. Based on sections of the landscape (Fig.\ref{fig:result3}F-J), we find that the potential landscape basins located in the spot state emerge suddenly at the beginning of the stripe/spot coexisting phase, whereas the ones located in the stripe state appear continuously at the beginning of the gap/stripe coexisting phase. Thus the critical transition from the stripe state to the spot state is discontinuous (the nucleation process), while the transition from the gap state to the stripe state is continuous or smooth. Additionally, both $J_{ave}$ and $e_p$ show a more pronounced decrease in the stripe/spot coexisting phase than in the gap/stripe coexisting phase. A sharper peak in EPR during a transition generally shows that this transition is more violent, leading to the same conclusions as those based on the landscape changes discussed above. 

In addition, to measure the time irreversibility of the system in practice from time series observations, we introduce the average differences $\overline{C}$ between two-point cross-correlations forward $C_1$ and backward $C_2$ in time (see SI for definitions). As shown in Fig.\ref{fig:result7}D, the obtained $\overline{C}$ of each pair of two modes and their summation show maximums in the coexisting phases, demonstrating that the time irreversibility of the system reaches its maximums during the transition process. And it is observed that $\overline{C}$ in the stripe/spot coexisting phase is much larger than the one in the gap/stripe coexisting phase, indicating that the time irreversibility in the former is larger than that in the later, further demonstrating that the transition from the stripe state to the spot state is more violent than the one from the gap state to the stripe state.

\section{Conclusion}

In summary, we developed the landscape and flux field theory combined with the mode expansion method under appropriate truncations to the spatial vegetation patterns in semi-arid ecosystems. The mode expansion method with appropriate truncations enabled us to reduce the substantial number of DOFs in spatial dynamics and identify several key spatial modes. As a consequence, the complex functional Fokker-Planck equation for the probabilistic evolution for stochastic spatial dynamics can be transformed into a more tractable Fokker-Planck equation in the mode space. Subsequently, by generalizing the landscape and flux theory to the spatial dynamics, not only the potential landscape of the ecosystem can be used to quantified the global stability via characterizing the states, their weights, and barrier height between states, but also the nonequilibrium flux field as well as the EPR can be obtained, which serves to reveal the dynamical and thermodynamical mechanisms of semi-arid ecosystems exhibiting different spatial patterns.

Our findings show that as $\beta$ increases, the landscape evolves successively from the potential basin located at gap to gap/stripe, stripe, stripe/spot, and finally spot spatial pattern states, consistent with the observed changes in vegetation patterns. Moreover, It was found that the nonequilibrium flux inside the landscape often points to the opposite direction of the potential gradient, acting as a driving force for the system to switch from one spatial pattern state to the another. The flux also contributes to the non-overlapping nature of transition paths between potential basins located at different spatial pattern states, demonstrating the time-reversal symmetry breaking of the ecosystem, further offering a new method to study the mechanism of pattern switching. Furthermore, both the averaged flux and EPR exhibit peaks near the phase boundaries, not only revealing the dynamical and thermodynamical mechanisms of the critical transition or bifurcation, but also serving as early warning signals for the desertification. Additionally, intensive simulations explored the influence of ecosystem size(noise), showing that as the size decreases(noise intensity increases), the critical transition shifts from a discontinuous to a continuous process, and may even disappear (see details in SI and Fig.S11). To verify the generality and robustness of our method, we also varied other parameters and observed similar results (see SI and Fig.S12,S13 for details).

The current study provides a convincing example to highlight the significant potential of utilizing the landscape and flux theory through the mode expansion method to nonequilibrium systems with spatial patterns formation and switching. Since spatial patterns exist widely in nature across physical, chemical, biological, ecological, and other fields, ranging from microscopic patterns in embryonic development to macroscopic patterns of vegetation or animal community, we believe that our work offers a significant and general approach to characterizing the pattern formation and switching in spatially extended systems. This not only provides insights into understanding the driving forces, but also allows us to reveal the underlying mechanisms of formation and switching of spatial patterns, such as the critical transition or bifurcation from one pattern state to another one. In addition, our method may provide a useful tool for practical applications in real systems, for example, giving an early warning of desertification or forest fire.

\section*{Acknowledgments}
J.S is grateful for support from NSFC12234019 and WIUCASQD2022012. W.W is thankful for support from NSFC12234019, WIUCASQD2023030 and WIUCASQD2022012 . S.A.L and D.P acknowledge support from NSF DMS Grant1951358.

\section*{AUTHORDECLARATIONS}
\subsection*{Competing interests}
The authors declare no competing interests.

\subsection *{Author Contributions}
J.W. and S.A.L. designed the research; J.S. and J.W. performed research; J.S., W.W., D.P., S.A.L. and J.W. contributed new analytic tools and analyzed data; All authors discussed the results and co-wrote the manuscript.

\section*{Data availability}
All study data are included in this article and/or SI Appendix.

%\bibliographystyle{unsrt}
%\bibliography{ref}

\end{document}